\documentclass[applemac]{aa}
\usepackage{natbib}
\usepackage{txfonts}
\usepackage{graphicx}
\usepackage{color}
\bibpunct{(}{)}{;}{a}{}{,}

\begin{document}

\title{Effect of grain size distribution and size-dependent grain heating on molecular abundances in starless and pre-stellar cores}
\author{O. Sipil\"a,
		B. Zhao,
		and P. Caselli
}
\institute{Max-Planck-Institute for Extraterrestrial Physics (MPE), Giessenbachstr. 1, 85748 Garching, Germany \\
e-mail: \texttt{osipila@mpe.mpg.de}
}

\date{Received / Accepted}

\abstract{We present a new gas-grain chemical model to constrain the effect of grain size distribution on molecular abundances in physical conditions corresponding to starless and pre-stellar cores. We introduce grain-size dependence simultaneously for cosmic-ray (CR)-induced desorption efficiency and for grain equilibrium temperatures. The latter are calculated with a radiative transfer code using custom dust models built for the present work. We keep explicit track of ice abundances on a set of grain populations. We find that the size-dependent CR desorption efficiency affects ice abundances in a highly non-trivial way that depends on the molecule. Species that originate in the gas phase, such as CO, follow a simple pattern where the ice abundance is highest on the smallest grains (which are the most abundant in the distribution). Some molecules, such as HCN, are instead concentrated on large grains throughout the time evolution, while others (like $\rm N_2$) are initially concentrated on large grains, but at late times on small grains, due to grain-size-dependent competition between desorption and hydrogenation. Most of the water ice is on small grains at high medium density ($n({\rm H_2}) \gtrsim 10^6 \, \rm cm^{-3}$), where the water ice fraction, with respect to the total water ice reservoir, can be as low as $\sim 10^{-3}$ on large (> 0.1 $\mu$m) grains. Allowing the grain equilibrium temperature to vary with grain size induces strong variations in relative ice abundances in low-density conditions where the interstellar radiation field and in particular its ultraviolet component are not attenuated. Our study implies consequences not only for the initial formation of ices preceding the starless core stage, but also for the relative ice abundances on the grain populations going into the protostellar stage. In particular, if the smallest grains can lose their mantles due to grain-grain collisions as the core is collapsing, the ice composition in the beginning of the protostellar stage could be very different than in the pre-collapse phase, owing to the fact that the ice composition depends strongly on the grain size.}


\titlerunning{Effect of grain size distribution on molecular abundances}
\maketitle

\section{Introduction}

Interstellar dust grains play a large role in the chemical and physical evolution of molecular clouds that host starless and pre-stellar cores -- the seeds of low-mass star formation. For example, the grains regulate the gas temperature in high-density regions where collisions between grains and gaseous molecules are frequent \citep[e.g.,][]{Goldsmith01}, and they also participate in the shaping of the chemical inventory in the clouds as new molecules are formed on their surface (see for example \citealt{Hama13} for a review).

The physical size of the dust grains varies from nanometers to some tenths of micrometers, or even larger depending on the environment. Various models for their size distribution and optical properties as a function of radius and bulk composition have been introduced in the literature \citep[e.g.,][]{Mathis77,Li01,Weingartner01,Jones13}. Yet in gas-grain chemical models used to predict molecular abundances in the interstellar matter (ISM), it almost ubiquitously assumed that the grains are monodisperse, that is, they all have the same radius, usually taken to be around 0.1\,$\mu$m \citep[e.g.,][]{Hasegawa92,Garrod06,Taquet12,Sipila15a,Vasyunin17}. This assumption greatly simplifies the chemical modeling, but it has negative consequences for the accuracy of the simulations. We highlight two issues here. First, grains are transiently heated to high temperatures by cosmic ray (CR) impacts which deposit energy into the grain; the amount of deposited energy depends, among other properties, on the physical size of the grain \citep{Fano63,Leger85}. Nanometer-sized grains can be heated to hundreds of\,K, while sub-micrometer grains are only heated to a few tens of~K \citep{Herbst06}. The transient heating allows efficient non-thermal desorption of molecules from the grain surface \citep{Hasegawa93a}. Second, the equilibrium temperature of a grain varies with its size. Considering that the efficiencies of chemical reactions on grain surfaces are strongly dependent on the grain temperature \citep[e.g.,][]{Hasegawa92}, coupled with the fact that non-thermal desorption allows the surface material to interact with the gas phase, it is highly desirable to investigate the effect that a size distribution may have on the chemical evolution of a star-forming object as a whole.

Indeed the effect of a grain size distribution on chemistry has been investigated in some previous studies, using models neglecting the effect of the size distribution on grain temperatures \citep{Acharyya11}, including size-dependent equilibrium temperatures \citep{Pauly16,Ge16}, and including grain-size-dependence in CR heating efficiency \citep{Zhao18,Iqbal18}. Of these studies, the work of \citet{Iqbal18} is the only one that considers size-dependent equilibrium temperatures and CR heating simultaneously, but their study was limited to a single set of physical conditions and lacked a treatment of size-dependence in grain cooling following a CR impact. Furthermore, the studies of \citet{Pauly16} and \citet{Ge16} considered a simple power-law dependence for the equilibrium dust temperature, or used temperature data appropriate for grains without ice mantles, respectively. Given that ice mantles start to form on grains already at low medium densities where one can reasonably expect size-dependent variations in grain equilibrium temperatures owing to the different response of the grain populations to heating by the interstellar radiation field (ISRF), there is a definite incentive to build a more complete model that ties the above together.

In this work, we present the first gas-grain chemical model that examines simultaneously the effects of grain-size-dependent CR desorption and grain equilibrium temperature on gas-phase and grain-surface abundances, applied to a variable set of physical conditions corresponding to those found in starless and pre-stellar cores. The heating and cooling timescales of the grains following CR impact events are scaled with the grain size, and the grain equilibrium temperatures are calculated with a radiative transfer code using dust models tailored for the present purposes. We track the ice abundances on each grain population explicitly in order to constrain the relative amounts of molecules as a function of grain size. We use the model to investigate ice abundances at the time of their formation in a low-density medium, and we also look into the relative ice abundances in high-density conditions preceding the protostellar stage.

The paper is structured as follows. Section~\ref{s:model} describes our model in detail. Section~\ref{s:results} presents the results of our model runs, applied to single-point physical conditions and to the pre-stellar core L1544. We discuss the results in Section~\ref{s:discussion}, where we also compare our work to the previous related studies available in the literature. Section~\ref{s:conclusions} presents our conclusions. Appendices~\ref{appendixA}~to~\ref{appendixC} present some additional results that complement the main text.

\section{Model}\label{s:model}

In this paper, we investigate the effect of a grain size distribution on molecular abundances in physical conditions corresponding to those found in starless and pre-stellar cores. Specifically, we study how the CR desorption process and the grain equilibrium temperature are affected by variations in grain size, and how these effects in turn influence gas-phase and grain-surface chemistry. This section presents the theoretical basis of our work, and details the chemical and physical models that we employ.

\subsection{Monodisperse grains versus grain size distribution}\label{ss:grainFormulae}

Gas-grain models of interstellar chemistry nearly universally assume monodisperse spherical dust grains, with the grain radius usually taken as 0.1\,$\mu$m. In such a case, the grain abundance with respect to hydrogen can be calculated from
\begin{equation}\label{eq:grainden}
X_{\rm d,m} = \frac{n_{\rm d}}{n_{\rm H}} = R_{\rm d} \frac{\mu m_{\rm H}}{\frac{4}{3} \pi a_{\rm m}^3 \rho_{\rm d}} \, ,
\end{equation}
where $n_{\rm H}$ is the total hydrogen number density ($\approx 2n({\rm H_2})$ inside dense clouds), $m_{\rm H}$ is the mass of the hydrogen atom, $\mu$ is the average particle mass, $\rho_{\rm d}$ is the grain material density, $a_{\rm m}$ is the monodisperse grain radius, and $R_{\rm d}$ is the dust-to-gas mass ratio. In this paper, we take $\mu = 1.4$, $\rho_{\rm d} = 2.5\,\rm g \, cm^{-3}$, and $R_{\rm d} = 0.01$.

If one considers instead a size distribution for the grains, a size scaling relation must be assumed. In this work, we adopt the grain size distribution from \citeauthor{Mathis77}\,(\citeyear{Mathis77}; hereafter MRN), where $dn_{\rm d}/da = C  n_{\rm H}  a^{-3.5}$. Here $C$ is a normalization constant whose value can be derived under the assumption of a constant dust-to-gas mass ratio:
\begin{equation}
C = 0.525 R_{\rm d} m_{\rm H} / (\pi \rho_{\rm d} (\sqrt{a_{\rm max}} - \sqrt{a_{\rm min}})) \, ,
\end{equation}
where $a_{\rm max}$ and $a_{\rm min}$ are respectively the maximum and minimum radii considered for the distribution. The MRN distribution is continuous, but it can be easily discretized for inclusion in a chemical model by dividing the size interval $\left[a_{\rm min},a_{\rm max}\right]$ into a number of grain size bins, calculating then the effective grain number density
\begin{equation}
X_{\rm d}^i = C \int_{a_{\rm min}^i}^{a_{\rm max}^i} a^{-3.5} \, da
\end{equation}
and the effective grain radius
\begin{equation}
a_{\rm eff}^i = \sqrt{\frac{\int a^2\,dn}{\int dn}} = \sqrt{\frac{\int_{a_{\rm min}^i}^{a_{\rm max}^i} a^{-1.5}\,da}{\int_{a_{\rm min}^i}^{a_{\rm max}^i} a^{-3.5}\,da}}
\end{equation}
in each size bin $i$. Finally, the effective grain area in each bin is given by
\begin{equation}
\left(\frac{n_{\rm d}\sigma}{n_{\rm H}}\right)_{\rm eff}^i = X_{\rm d}^i \pi (a_{\rm eff}^i)^2 \, .
\end{equation}

Presently we are interested in a comparison between models where we assume either monodisperse grains or a grain size distribution. These are two inherently different cases. To facilitate the comparison, we set the monodisperse grain radius by requiring that the resulting total grain surface area ($X_{\rm d,m}\pi a_{\rm m}^2$) matches the total effective surface area of the size distribution ($\Sigma_i \left( n_{\rm d}\sigma / n_{\rm H}\right)_{\rm eff}^i$). In this paper we assume $\left[a_{\rm min},a_{\rm max}\right] = \rm \left[0.03\,\mu m,0.3\,\mu m\right]$ for the size distribution limits, leading to a monodisperse grain radius of $a_{\rm m} = 0.095$\,$\mu \rm m$ based on the area constraint. The grains in the size distribution model are always assumed to be spherical.

Equalizing the total grain surface area between the two models means that the (total) grain abundance is not the same in both cases ($X_{\rm d,m} \neq \Sigma_i X_{\rm d}^i$). An alternative approach would be to set the monodisperse grain radius so that the grain abundance equals the total abundance in the size distribution model, but we consider this approach inferior as it would lead to a discrepancy in the total grain surface area, which is of critical importance to the present study as adsorption rates depend on the grain area. We note that the dust-to-gas mass ratio is never varied from its fiducial value of~0.01, meaning that the size distribution and monodisperse grain models correspond to the same total dust mass.

\begin{table*}
        \renewcommand{\arraystretch}{1.3}
        \centering
        \caption{Parameters related to the grain size distribution discussed in the main text.}
        \begin{tabular}{c | c | c | c | c | c}
                \hline
                \hline
                Size bin label & 1 &  2 & 3 & 4 & 5 \\
                \hline
                $a_{\rm min}, a_{\rm max} \, [{\rm \mu m}]$ & 0.030, 0.048 &  0.048, 0.075 & 0.075, 0.119 & 0.119, 0.189 & 0.189, 0.300 \\
                \hline
                Effective radius$ \, [{\rm \mu m}]$ & 0.037 & $0.058$ & $0.092$ & $0.146$ & $0.232$\\
                \hline
                Abundance$^{(a)}$ & $4.37 \times 10^{-12}$ & $1.38 \times 10^{-12}$ & $4.37 \times 10^{-13}$ & $1.38 \times 10^{-13}$ & $4.37 \times 10^{-14}$ \\
                \hline
                $T_{\rm max} \, [{\rm K}]$ & $162.1$ & $109.5$ & $75.4$ & $53.6$ & $39.5$\\
                \hline
                $R_{\rm cool} \, [{\rm s}]$ & $5.90 \times 10^{-10}$ & $2.06 \times 10^{-8}$ & $2.96 \times 10^{-6}$ & $1.89 \times 10^{-3}$ & $5.42$\\
                \hline
                $R_{\rm heat} \, [{\rm s}]$ & $2.33 \times 10^{14}$ & $9.29 \times 10^{13}$ & $3.70 \times 10^{13}$ & $1.47 \times 10^{13}$ & $5.86 \times 10^{12}$\\
                \hline
                $f(a,T_{\rm max})$ & $2.53 \times 10^{-24}$ & $2.21 \times 10^{-22}$ & $8.00 \times 10^{-20}$ & $1.28 \times 10^{-16}$ & $9.24 \times 10^{-13}$\\
                \hline
                $k_{\rm CR}^{\rm CO} \, [{\rm s^{-1}}]$ & $2.13 \times 10^{-15}$ & $6.21 \times 10^{-15}$ & $1.92 \times 10^{-14}$ & $6.31 \times 10^{-14}$ & $2.21 \times 10^{-13}$\\
                \hline
        \end{tabular}
        \label{tab:grainParameters}
        \tablefoot{$^{(a)}$ With respect to $n_{\rm H}$.}
\end{table*}

\subsection{CR-induced desorption}

\citet{Hasegawa93a} derived an expression for the rate coefficient of a CR-induced desorption (hereafter simply ``CR desorption'') event, which for the desorption of a grain-surface species $j$ takes the form
\begin{equation}\label{CRdesorption}
k_{\rm CR}^j = f(70\,{\rm K}) k_{\rm therm}^j(70\,{\rm K}) \, ,
\end{equation}
where $k_{\rm therm}$ is the thermal evaporation rate coefficient at 70\,K. This expression is based on the argument that energetic iron nuclei deposit enough energy when passing through a grain with radius 0.1\,$\mu$m to heat it to 70\,K, and that desorption effectively takes place at 70\,K; the molecular evaporation cools the grain back to 10\,K. The $f(70\,{\rm K})$ efficiency term is the grain ``duty cycle'' at 70\,K, defined as the ratio of the grain cooling timescale ($R_{\rm cool} \sim 10^{-5} \, \rm s$) to the time interval between successive CR heating events for 0.1\,$\mu$m grains, which \citet{Hasegawa93a} calculated to be $R_{\rm heat} = 3.16 \times 10^{13} \, \rm s$ based on the iron CR flux given by \citet{Leger85}. The heating event interval scales as $a^{-2}$ \citep{Leger85}, while the cooling timescale is independent of grain radius; the latter scales with the temperature as $\exp(-E_{\rm b}/T)$, where $E_{\rm b}$ is the binding energy of a typical volatile grain-surface species, which \citet{Hasegawa93a} fixed to 1200\,K. We thus obtain for the grain-size-dependent duty cycle:
\begin{equation}
f(a,T_{\rm max}) = f(0.1\,{\rm \mu m}, 70\,{\rm K}) \times \frac{ e^{{\rm 1200\,\rm K}/{T_{\rm max}}} }{ e^{{\rm 1200\,\rm K}/{\rm 70\,\rm K}} }\times \left( \frac{a}{0.1\,{\rm \mu m}} \right)^2 \, ,
\end{equation}
where $T_{\rm max}$ is the radius-dependent transient maximum temperature that the grain reaches upon a CR impact, $f(0.1\,{\rm \mu m}, 70\,{\rm K}) = 3.16 \times 10^{-19}$ is the duty cycle for 0.1\,$\mu$m grains \citep{Hasegawa93a}, and the last two terms represent the scaling of $R_{\rm cool}$ and $R_{\rm heat}$, respectively. $T_{\rm max}$ is derived from the grain heat capacity \citep{Leger85}. Small grains will be transiently heated to temperatures in excess of 150\,K (the upper limit in \citealt{Leger85}); here we use the extrapolated heat capacity from \citet{Zhao18} to calculate $T_{\rm max}$ for the smallest grains in the distribution.

We divided the grain size distribution into five logaritmically spaced size bins, and derived the effective grain radius and abundance in each bin using the formulae presented in Sect.\,\ref{ss:grainFormulae}. The transient maximum temperatures, successive heating intervals and duty cycles were then calculated based on the effective grain radii. The results of these calculations are collected in Table~\ref{tab:grainParameters}, where we also show for reference the CR desorption rate coefficient of CO, taking 1150\,K for the CO binding energy. CR desorption rate coefficients increase with grain radius. The duty cycle $f(a,T_{\rm max})$ varies by more than ten orders of magnitude over the size distribution and obtains a very low value for the smallest grains, but the CR desorption rate coefficient varies only by approximately two orders of magnitude over the distribution; this is because the low duty cycle values are compensated by the high transient maximum temperatures for the smallest grains. The number of grain size bins has an effect on our results, but only for the relative ice abundances in the various grain size bins and not for the total ice abundances; we discuss this issue in more detail in Appendix~\ref{appendixA}.

\subsection{Size-dependent equilibrium grain temperature}\label{ss:sizeDependentTemperatures}

The grain size affects the equilibrium temperature ($T_{\rm dust}$) of the grains. Chemical reactions on grain surfaces are very sensitive to variations in $T_{\rm dust}$ owing to two factors: 1) The thermal diffusion rate on the surface scales as $\exp\left(-E_{\rm b}^j/T_{\rm dust}\right)$, where $E_{\rm b}^j$ is the binding energy of species $j$, and 2) Barrier-mediated surface reactions scale analogously by the inverse of $T_{\rm dust}$. Therefore, allowing the equilibrium grain temperature to vary over the size distribution will affect the reactivity on the different grain populations. For this reason we track explicitly the chemical abundances on the grains in each size bin -- the chemical model thus contains five versions of all surface species in our current setup of five grain size bins.

In several previous works concerning models of the structure of starless and prestellar cores \citep[e.g.,][]{Sipila17a}, we have used opacity data from \citet{Ossenkopf94} to derive $T_{\rm dust}$ profiles as a function of density. In the present work, we cannot employ the same approach because the \citet{Ossenkopf94} tables are derived over a broad range of grain sizes (from $5\times10^{-3}$\,$\mu$m to 0.25\,$\mu$m), and we are presently considering a setup with multiple grain size bins. We therefore generated new sets of absorption and scattering cross sections for the five size bins displayed in Table~\ref{tab:grainParameters}, and over a broad distribution from 0.03 to 0.3\,$\mu$m, using the SIGMA code (Lef\`evre et al.\,(subm.)\footnote{https://github.com/charlenelefevre/SIGMA}; \citealt{Woitke16}; \citealt{Min05}). We assumed that the grains consist of a carbon/silicate core in a 50/50 volume ratio (matching the grain material density of $\rho_{\rm d} = 2.5\,\rm g \, cm^{-3}$ in the chemical code) and are coated with a water ice layer with a volume fraction half that of the core. The refractive index data is taken from \citet{Ossenkopf94} (included in SIGMA) so that this setup corresponds roughly to the ``thin ices'' model of \citet{Ossenkopf94}. Radiative transfer calculations to determine $T_{\rm dust}$ profiles were carried out using the CRT code \citep{Juvela05}; for these, we assumed that the core is surrounded by an external layer corresponding to a visual extinction of $A_{\rm V} = 1 \, \rm mag$, and we took the ISRF spectrum from \citet{Black94}. The results of these calculations are presented in Sect.\,\ref{ss:L1544}.

\begin{table*}
        \renewcommand{\arraystretch}{1.3}
        \centering
        \caption{Model cases considered in this paper.}
        \begin{tabular}{c | c }
                \hline
                \hline
                Model denomination & Description \\
                \hline
                M1 & Monodisperse grains with $a = 0.095\,\mu \rm m$ (see Sect.\,\ref{ss:grainFormulae} for details) \\ 
                \hline
                M2 & Grain size distribution with radii and abundances given in Table~\ref{tab:grainParameters}, but $T_{\rm max}$ and duty cycle\\
                & set to constant values for all grain sizes (see main text) \\ 
                \hline
                M3 & Grain size distribution with parameters given in Table~\ref{tab:grainParameters} \\ 
                \hline
                M4 & As M3, but equilibrium $T_{\rm dust}$ depends on grain size\\ 
                \hline
        \end{tabular}
        \label{tab:modelCases}
\end{table*}

\begin{figure*}
	\includegraphics[width=2.0\columnwidth]{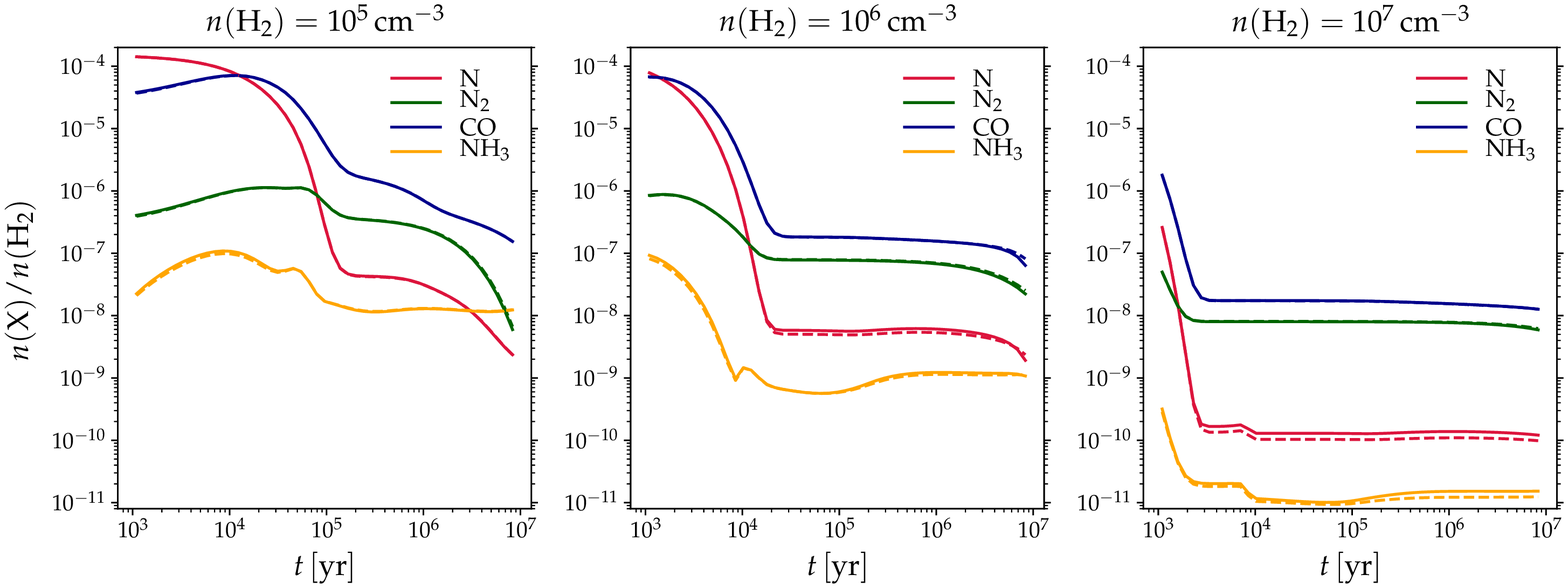}
    \caption{Abundances of selected gas-phase species, indicated in the figure, as a function of time for three medium densities. Solid lines correspond to model~M1, while dashed lines correspond to model~M2.}
    \label{fig:M1vsM2}
\end{figure*}

\subsection{Additional model details}\label{ss:modelDetails}

The results presented in this paper pertain to four individual model cases. These are collected in Table~\ref{tab:modelCases}. Model~M1 represents the standard case of monodisperse grains, which is almost universally adopted in gas-grain models of ISM chemistry. In model~M2, we consider instead a grain size distribution, assuming that $T_{\rm max}$ and the duty cycle are constant for all grain sizes. This model is included specifically to illustrate the effect of ignoring the scaling of the CR desorption rate coefficients with grain radius and temperature. Here, $T_{\rm max,m} = 73.86\, \rm K$ and $f(0.095\,\mu{\rm m},T_{\rm max,m}) = 1.16 \times 10^{-19}$ are respectively the temperature and duty cycle that one obtains for the monodisperse grains in model~M1, so that models M1 and M2 are comparable. Model~M3 represents the fully radius-dependent distribution model as detailed above (cf. Table~\ref{tab:grainParameters}). Model~M4 introduces into M3 the grain-size-variable $T_{\rm dust}$ calculated using CRT.

We use the gas-grain chemical code discussed in \citeauthor{Sipila19a}\,(\citeyear{Sipila19a}; see also references therein), updated to accommodate for the grain size distribution. In this paper we employ point-like physical models to demonstrate the effects on the chemistry due to different grain setups. For simplicity we adopt in these models a constant temperature $T_{\rm gas} = T_{\rm dust} = 10\,\rm K$ (that is, model~M4 is not considered), and a constant visual extinction $A_{\rm V} = 10\,\rm mag$. The density of the medium is varied between $n({\rm H_2}) = 10^5\,\rm cm^{-3}$ and $10^7\,\rm cm^{-3}$.

\begin{table}
	\centering
	\caption{Initial abundances (with respect to $n_{\rm H}$) used in the chemical modeling.}
	\begin{tabular}{l|l}
		\hline
		\hline
		Species & Abundance\\
		\hline
		$\rm H_2$ & $5.00\times10^{-1}$\\
		$\rm He$ & $9.00\times10^{-2}$\\
		$\rm C^+$ & $7.30\times10^{-5}$\\
		$\rm N$ & $5.30\times10^{-5}$\\
		$\rm O$ & $1.76\times10^{-4}$\\
		$\rm S^+$ & $8.00\times10^{-8}$\\
		$\rm Si^+$ & $8.00\times10^{-9}$\\
		$\rm Na^+$ & $2.00\times10^{-9}$\\
		$\rm Mg^+$ & $7.00\times10^{-9}$\\
		$\rm Fe^+$ & $3.00\times10^{-9}$\\
		$\rm P^+$ & $2.00\times10^{-10}$\\
		$\rm Cl^+$ & $1.00\times10^{-9}$\\
		\hline
	\end{tabular}
	\label{tab:initialabundances}
\end{table}

\begin{figure*}
	\includegraphics[width=2.0\columnwidth]{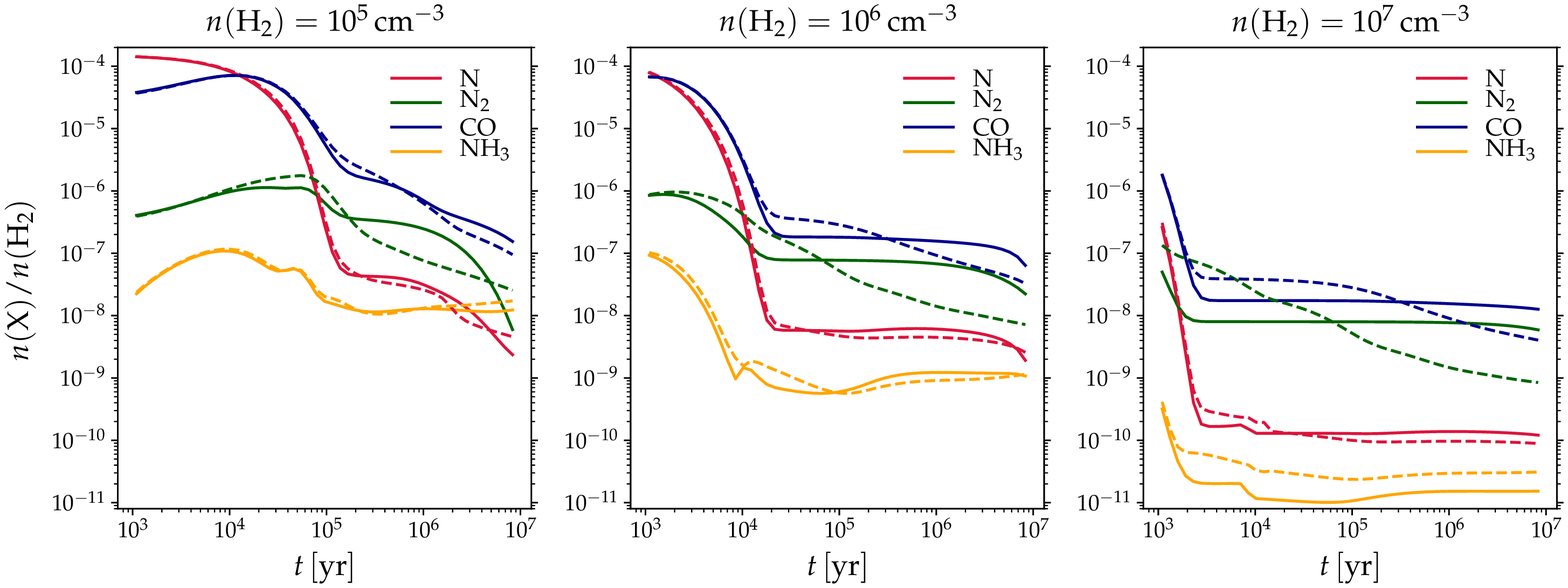}
    \caption{As Fig.\,\ref{fig:M1vsM2}, but dashed lines correspond to model~M3.}
    \label{fig:M1vsM3}
\end{figure*}

In \citet{Sipila19a}, we presented the results of an extensive benchmarking study attempting to reproduce the observed distribution of ammonia in the pre-stellar core L1544, for which we used the physical model by \citet{Keto10a}. We construct a similar model here for the purposes of studying the size distribution effects in the context of a real core that has a non-constant density and temperature structure. We derive time-dependent chemical abundance profiles in L1544 by using dust temperature profiles designed specifically for the present work (see Sect.\,\ref{ss:L1544}) in connection with the density and gas temperature structure from \citet{Keto10a}. We divide the so-obtained physical source model into concentric shells and calculate chemical evolution separately in each shell. This process of deriving time-dependent abundance profiles is described in more detail in \citet{Sipila19a}; we employ here the same initial chemical abundances (Table~\ref{tab:initialabundances}) and the same chemical networks as in that paper\footnote{Although our model includes deuterium and spin-state chemistry, we investigate neither deuterated species nor spin chemistry in the present work. The water abundances shown in the figures in this paper represent a sum over the ortho and para forms.}. An $A_{\rm V}$ profile, important for the photoreactions in the chemical model, was constructed assuming an external visual extinction of $A_{\rm V}^{\rm ext} = 1\,\rm mag$ (matching the radiative calculations). We stress that the L1544 core model is adopted just so we can explore the influence of the grain size distribution on chemical evolution using a core model with a high central density -- here we do not attempt to reproduce any observations toward L1544, and our present aims could be fulfilled using any generic core model, such as a Bonnor-Ebert sphere.

\section{Results}\label{s:results}

\subsection{Effect of grain size distribution and size-dependent CR desorption on molecular abundances}

We proceed with a step-by-step analysis to uncover the effect that the grain size distribution has on molecular abundances in dense core conditions. Our starting point is the comparison of models~M1 and M2; the two are equivalent except of the introduction of a distribution of grain sizes in the latter model. Figure~\ref{fig:M1vsM2} shows the abundances of some selected gas-phase species as a function of time at three $\rm H_2$ densities. The plot demonstrates clearly that simply introducing a distribution of grain sizes, and keeping the CR desorption scheme unaltered, has no significant effect on the abundances of the various species. This conclusion holds for all species in our model, and thus is not limited to the species that we selected for Fig.\,\ref{fig:M1vsM2}, although the move to a distribution of grain sizes naturally means that grain-surface species are spread out over several populations of grains (see below).

\begin{figure*}
	\includegraphics[width=2.0\columnwidth]{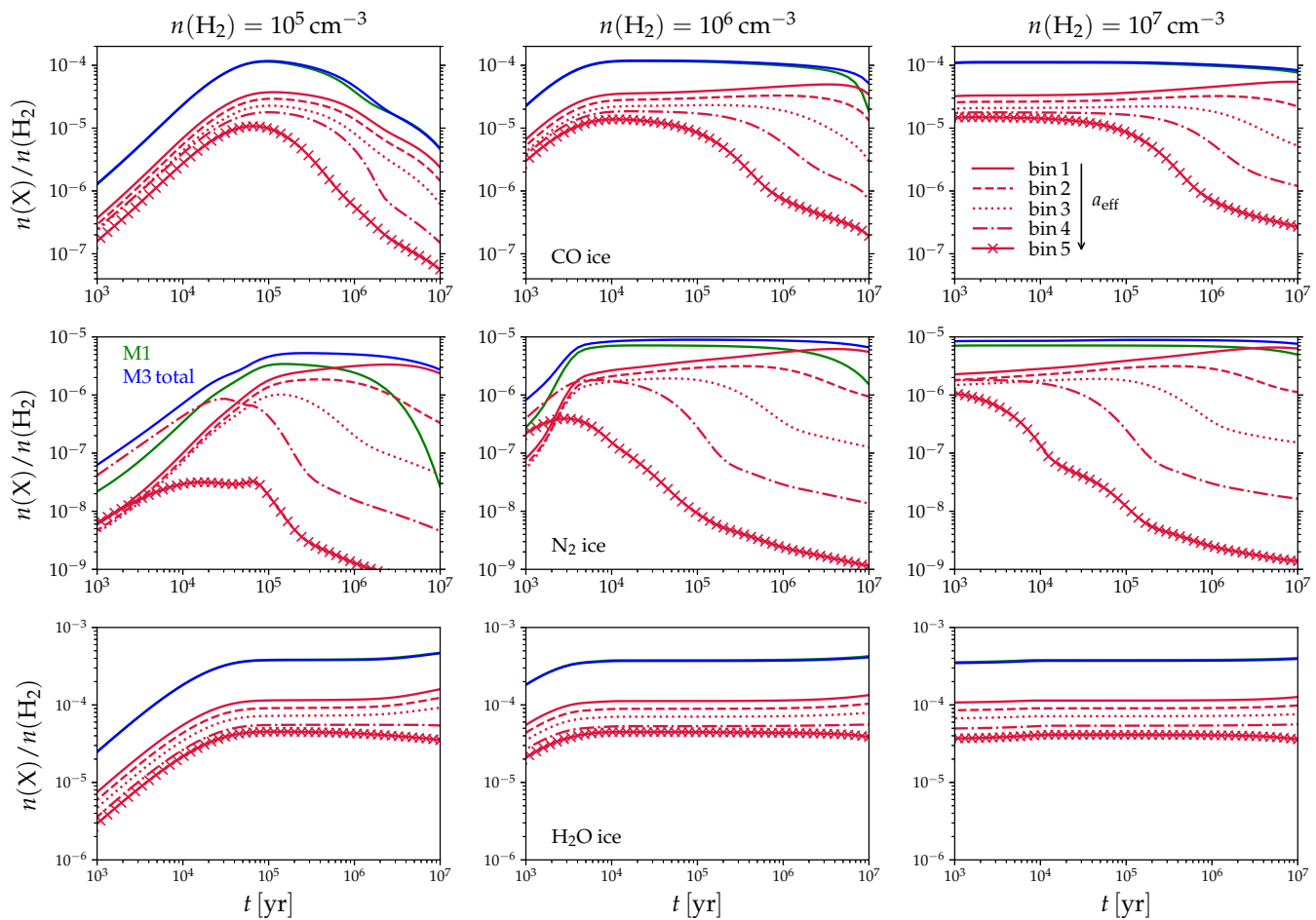}
    \caption{Distributions of CO ({\sl top row}), $\rm N_2$ ({\sl middle row}), and $\rm H_2O$ ({\sl bottom row}) ice as functions of time at three different medium densities. Green solid lines correspond to model~M1. Red and blue lines correspond to model~M3; the red lines represent the abundances on each grain population, while the blue line represents the sum over the individual grain populations. The green and blue lines overlap almost perfectly for $\rm H_2O$ ice. The effective grain radius increases with bin label as indicated in the top right panel.}
    \label{fig:M1vsM3_grains}
\end{figure*}

The effect of the grain size distribution becomes evident as we introduce the more realistic case of grain-size-dependent transient maximum temperatures and duty cycles (model~M3), although the magnitude of the effect varies with each species. Fig.\,\ref{fig:M1vsM3} shows a comparison of models~M1~and~M3. The size distribution model predicts an increase in abundances for the various species until $t \sim 10^5 \, \rm yr$, compared to the monodisperse grains model, with an increasing effect toward higher densities. At longer timescales the situation is -- in general -- reversed. These differences can be understood by careful examination of grain-surface molecular distributions and their effect on the gas phase. The surface distributions can be highly non-trivial when the grain size is allowed to vary, depending on the species and on the time.

We show in Fig.\,\ref{fig:M1vsM3_grains} the abundances of CO, $\rm N_2$, and $\rm H_2O$ ice as functions of time. Let us consider $\rm N_2$ in the lowest-density model ($n({\rm H_2}) = 10^5 \, \rm cm^{-3}$) at $t = 10^4 \, \rm yr$. At this time, gas-phase $\rm N_2$ is formed mainly through the $\rm N + CN \longrightarrow C + N_2$ reaction in both cases (M1~and~M3). However, the contribution of CR-desorbed icy $\rm N_2$ to the gas-phase $\rm N_2$ abundance is only less than 1\% in model~M1, but about 10\% in model~M3. This accounts for the slightly higher gas-phase $\rm N_2$ abundance in the latter model. $\rm N_2$ formation in the ice competes against hydrogenation and desorption of atomic~N. Hydrogenation is the most efficient destruction process of atomic~N on all grains, but its relative efficiency diminishes toward larger grains because of the associated increase of CR desorption rate coefficients (Table~\ref{tab:grainParameters}). As the grain size increases, atomic~H is desorbed more effectively, boosting $\rm N_2$ production through the $\rm N + N$ association reaction. However, for the largest grains, CR desorption of atomic~N is efficient enough to start inhibiting $\rm N_2$ formation. Therefore, there exists a ``sweet spot'' for grain-surface $\rm N_2$ production, which in our model occurs for grains slightly larger than 0.1 micron. This is clearly evident in Fig.\,\ref{fig:M1vsM3_grains}, where grain-surface $\rm N_2$ is in model~M3 mainly on the grains in the fourth size bin ($a_{\rm eff} = 0.146 \, \rm \mu m$) at $t = 10^4 \, \rm yr$. CR desorption rate coefficients in model~M1 are smaller than those on the largest grains in model~M3, and so the total effect of CR desorption is larger in the latter model at early times. As the medium density is increased, CR desorption events occur more often, and gas-phase $\rm N_2$ formation is dominated by CR desorption in both models~M1~and~M3; the $\rm N_2$ ice abundance is highest on the largest grains only for a period of a few hundred years at $n({\rm H_2}) = 10^7 \, \rm cm^{-3}$.

At later times, between $t \sim 10^5 \, \rm yr$ and a few $\times$ $10^6 \, \rm yr$ in the model with $n({\rm H_2}) = 10^5 \, \rm cm^{-3}$, model~M3 predicts a lower gas-phase $\rm N_2$ abundance (Fig.\,\ref{fig:M1vsM3}), even though $\rm N_2$ ice is more abundant than in model~M1. The difference is again directly related to grain chemistry. The main formation pathway for gas-phase $\rm N_2$ at late times is $\rm N_2H^+ + e^- \longrightarrow N_2 + H$, and its main destruction pathway is $\rm H_3^+ + N_2 \longrightarrow N_2H^+ + H_2$, constituting a loop between $\rm N_2$ and $\rm N_2H^+$. This applies to both models M1~and~M3, and in fact the abundance of $\rm H_3^+$ is, at late times, somewhat higher in model~M3 than in M1. Inspection of Fig.\,\ref{fig:M1vsM3_grains} reveals that $\rm N_2$ ice is abundant only on  the smallest grains at $t > 10^5 \, \rm yr$ because of the high CR desorption efficiency on large grains, and so the net CR desorption rates are at late times higher in model~M1 than in model~M3.

The distribution of CO ice is straightforward: the larger the grain, the lower the CO ice abundance. CO is formed mainly in the gas phase, and the majority of CO ice is material deposited from the gas phase through adsorption (unlike $\rm N_2$ ice which is formed {\sl in situ} at early times). Adsorption rates are largest on small grains which have the greatest surface area, and hence the CO ice abundance simply decreases with increasing grain radius. After $t \sim 10^5 \, \rm yr$ when CR desorption becomes important, larger abundance variations between grains of different radius develop due to the increase of the CR desorption efficiency toward larger grains (Table~\ref{tab:grainParameters}). CO is not affected by this effect as greatly as $\rm N_2$, owing to its higher binding energy. The total CO ice abundance profile displays a clear peak at $n({\rm H_2}) = 10^5\,\rm cm^{-3}$, while remaining flat throughout the time evolution at $n({\rm H_2}) = 10^7\,\rm cm^{-3}$. The efficiency of CO ice processing is tied to the abundance of gas-phase atomic~H, which is roughly inversely proportional to the medium density. At $n({\rm H_2}) = 10^5\,\rm cm^{-3}$, the total CO ice abundance decreases past $\sim10^5$\,yr due to hydrogenation processes that convert CO into formaldehyde and further into methanol. The production of methane is also effective. At higher densities, less H atoms are available for hydrogenation, and the decline in CO abundance occurs only to a rather limited degree; the abundances of formaldehyde and methanol ice are correspondingly much lower than at $n({\rm H_2}) = 10^5\,\rm cm^{-3}$.

Much like that of CO, the distribution of $\rm H_2O$ ice is in these physical conditions a simple decreasing function of the grain size label. However, because of the high binding energy of water ice (5700\,K in our model), no significant time-dependence is exhibited in the abundance ratios between ices on the different grain populations.

Figure~\ref{fig:M1vsM3} shows that the gas-phase ammonia abundance is slightly boosted by the introduction of a grain size distribution at high density, and does not experience a similar drop with time with respect to model~M1 as the CO and $\rm N_2$ abundances do. In general, how the abundance of a given species responds to alterations in the grain model is dependent on several factors, such as whether that species is formed mostly in the gas phase or on grains, on the volatility of the species, etc. (We return to this issue in Sect.\,\ref{ss:otherAbundances}.) Nevertheless it is clear already from the very small sample of species displayed in the preceding figures that switching from monodisperse grains to a size distribution has a marked effect on the overall chemistry, especially at high density. However, we note that here we started from initially atomic gas (except for hydrogen and deuterium which are initially in $\rm H_2$ and HD) in order to better bring out the density dependence in the effect that the different grain models have on the chemistry. This of course represents an extreme case, and in reality one expects the gas density to increase gradually as the pre-stellar core forms and ultimately collapses, which may mean that differences between models M1~and~M3 will not be as great as those displayed in Fig.\,\ref{fig:M1vsM3}.

\subsection{Grain-size-dependent equilibrium dust temperature and molecular abundances in L1544}\label{ss:L1544}

\begin{figure}
	\includegraphics[width=1.0\columnwidth]{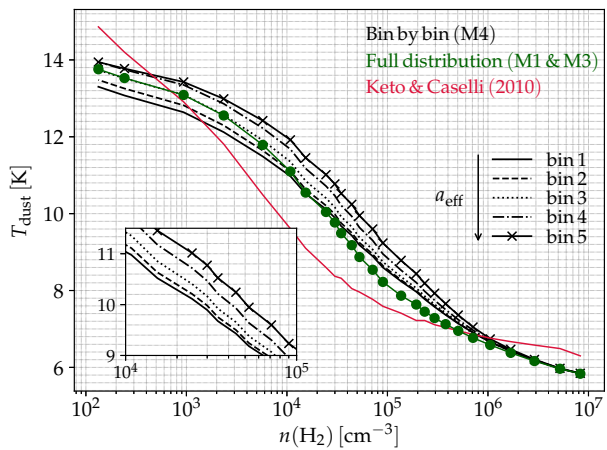}
    \caption{Equilibrium dust temperature profiles as a function of density, calculated for the five distinct grain populations (black lines; model~M4) or over the entire distribution (green; models~M1~and~M3). The effective grain radius increases with bin label as indicated in the plot. The red line shows the $T_{\rm dust}$ profile from \citet{Keto10a}. The inset shows a zoom-in of the $n({\rm H_2}) = 10^4\,\rm cm^{-3}$ to $n({\rm H_2}) = 10^5\,\rm cm^{-3}$ density regime, excluding the result for the full distribution.}
    \label{fig:L1544_temperatures}
\end{figure}

\begin{figure*}
	\includegraphics[width=2.0\columnwidth]{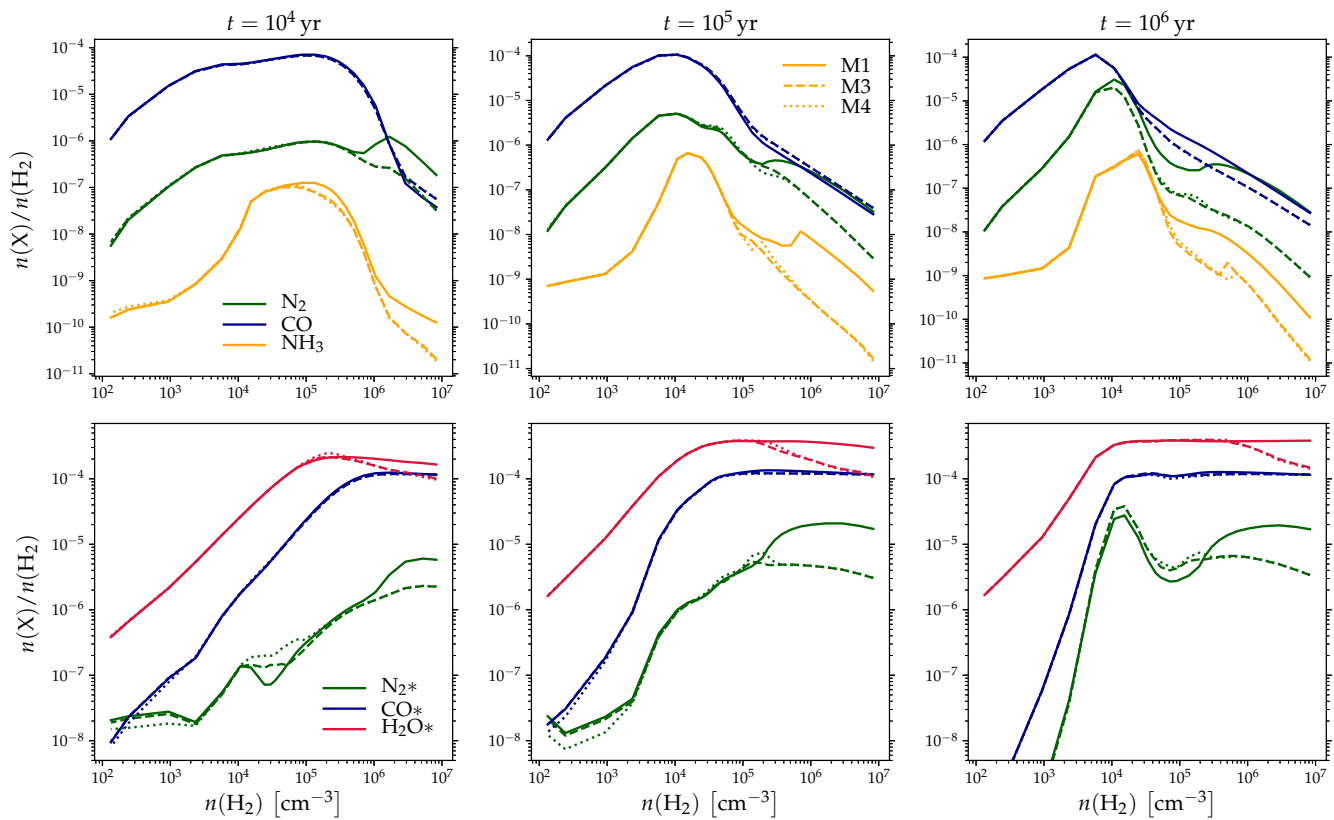}
    \caption{Abundances of selected gas-phase species ({\sl top row}) and grain-surface species (labeled with asterisks; {\sl bottom row}) as functions of medium density in the L1544 model. The abundances are displayed at three distinct time steps, ranging from $10^4\,\rm yr$ ({\sl left}) to $10^6\,\rm yr$ ({\sl right}). The different linestyles correspond to model~M1 (solid lines), model~M3 (dashed lines), and model~M4 (dotted lines). The ice abundances represent sums over all populations in models~M3~and~M4.}
    \label{fig:L1544_density}
\end{figure*}

Next, we examine the differences in the predictions of models~M1, M3, and M4 when applied to the pre-stellar core L1544. Figure~\ref{fig:L1544_temperatures} presents $T_{\rm dust}$ profiles obtained using CRT, when applying the dust model discussed in Sect.\,\ref{ss:sizeDependentTemperatures} to the L1544 density structure. We also show the $T_{\rm dust}$ profile of \citet{Keto10a} for reference. The $T_{\rm dust}$ profile calculated over the full grain size distribution is used in models~M1~and~M3. In the case of model~M4, the equilibrium $T_{\rm dust}$ increases with grain size, which holds throughout the range of densities in the core. The difference between the temperatures of the smallest and largest grains ($\Delta T$) is the largest between $n({\rm H_2}) = 10^4\,\rm cm^{-3}$ and $n({\rm H_2}) = 10^5\,\rm cm^{-3}$, where $\Delta T \sim 1 \, \rm K$, and in general $\Delta T$ is close to 1\,K in a broad density range from the lowest densities up to $\sim 10^5 \, \rm cm^{-3}$. The low end of this range represents the medium densities where ices form in the first place, while at the high end molecular freeze-out becomes important. The $T_{\rm dust}$ profile calculated over the entire distribution is largely similar to those of the individual grain populations, but presents a somewhat different slope in the $n({\rm H_2}) = 10^4\,\rm cm^{-3}$ to $n({\rm H_2}) = 10^6\,\rm cm^{-3}$ density range. We note that the dust temperature profiles calculated here are clearly different from the one obtained by \citet{Keto10a}. The new dust model constructed for the present work is not meant to reproduce observed temperatures or (submillimeter) intensities toward L1544, but is merely designed to investigate the effect of variations in equilibrium $T_{\rm dust}$ on ice abundances.

A gas temperature profile must be supplied in order to carry out chemical calculations using the L1544 density structure. The gas temperature affects grain-surface chemistry through its influence on adsorption rates, which depend on the thermal speed of the molecules, and it naturally impacts the synthesis of molecules in the gas phase. Our aim is to investigate the effect that the different approaches to the treatment of grain heating have on surface chemistry, and thus for models~M1 to M4 to be maximally comparable, we adopt for simplicity the gas temperature profile of \cite{Keto10a} in all cases. It would be possible to calculate new gas temperature profiles for the present purposes using for example the hydrodynamics code presented in \citet{Sipila18} and \citet{Sipila19a}, but because the gas temperature also depends on the dust temperature through the dust-gas collisional coupling, the gas temperature profiles in models~M3~and~M4 would not be equal. Moreover, carrying out this calculation is not straightforward for model~M4 where there are several dust temperature components to consider. We thus retain the \cite{Keto10a} gas temperature profile for L1544 so that the effect of varying dust temperature can be more easily distinguished between the models.

\begin{figure*}
\centering
	\includegraphics[width=2.0\columnwidth]{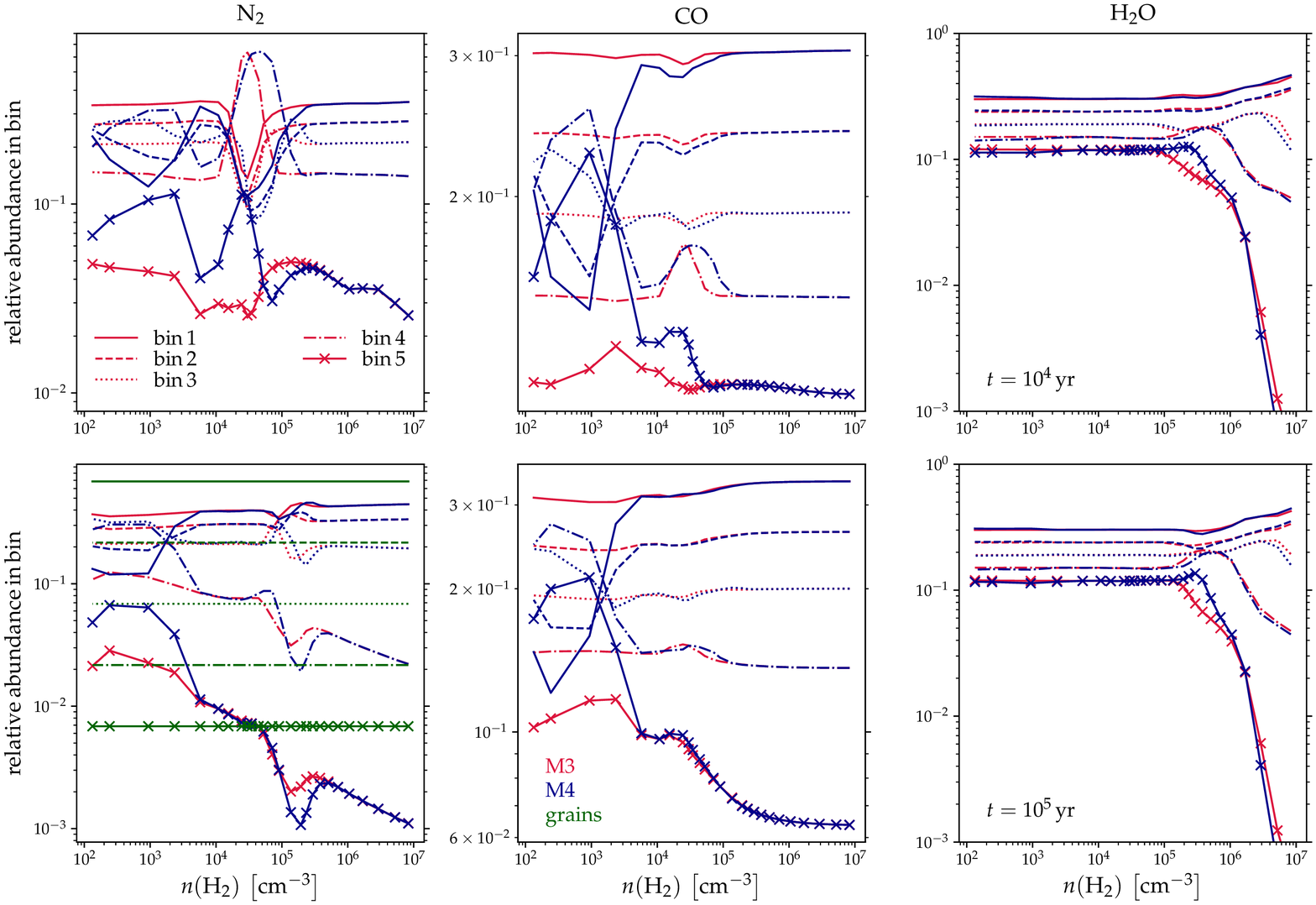}
    \caption{Relative abundances in each grain size bin as functions of density in the L1544 model at $t = 10^4 \, \rm yr$ ({\sl top}) and at $t = 10^5 \, \rm yr$ ({\sl bottom}). Red lines correspond to model~M3, while blue lines correspond to model~M4. Green lines display the relative abundance of the grains in each bin, shown only in the bottom-left panel for clarity (but note that the y-axis scaling is different in each subplot). The effective grain radius increases with bin label.}
    \label{fig:L1544_density_ices}
\end{figure*}

Figure~\ref{fig:L1544_density} shows selected molecular abundance profiles in the L1544 model at different durations of chemical evolution starting from the (mostly) atomic initial state. The gas-phase $\rm N_2$ and CO abundances follow the trends displayed in Fig.\,\ref{fig:M1vsM3}: $\rm N_2$ is strongly depleted at high densities ($n({\rm H_2}) \gtrsim 10^5\,\rm cm^{-3}$ in models~M3~and~M4 from $\sim 10^5\,\rm yr$ onwards), while the CO abundance takes a somewhat longer time to drop below the model~M1 value. There is no significant difference between models~M3~and~M4 for the gas-phase species.

Deviation from the results displayed in Fig.\,\ref{fig:M1vsM3_grains} is evident for $\rm N_2$ ice; the abundance is a factor of 2-3 lower in models~M3~and~M4 than in model~M1. Water ice presents similar behavior. This effect is tied to the response of grain chemistry to the temperature ($\sim$6\,K versus 10\,K in Fig.\,\ref{fig:M1vsM3_grains}). Decreasing the temperature allows atomic hydrogen to subsist longer on the grain surfaces, increasing hydrogenation rates. There is a consequent overall boost to the abundances of, for example, nitrogen hydride ices when the temperature is decreased from 10 to 6\,K. The difference between models~M1~and the two size distribution models is due to grain-size-dependent CR desorption efficiency. In models~M3~and~M4 where the CR desorption rates depend on the grain size, the (average) CR desorption rate of atomic hydrogen is lower than in model~M1, leading to higher nitrogen hydride ice abundances and hence a lower $\rm N_2$ ice abundance in the size distribution models. The gas-phase abundance of ammonia decreases with the more efficient destruction of $\rm N_2$ ice in models~M3~and~M4 (Fig.\,\ref{fig:L1544_density}). The response of the models to changes in $T_{\rm dust}$ is highly non-trivial and highlights the need for precise dust temperature modeling as opposed to assuming that the temperature is constant (typically 10\,K) in the shielded high-density regions inside pre-stellar cores.

Figure~\ref{fig:L1544_density_ices} presents a breakdown of the abundances of CO, $\rm N_2$, and $\rm H_2O$ ice as functions of medium density at $t = 10^4 \, \rm yr$ and $t = 10^5 \, \rm yr$ in models~M3~and~M4. Here, we show the relative ice abundance in each grain size bin, that is, the abundance in each bin divided by the total abundance (sum over all bins). We also show in one panel the relative abundances of each grain population for reference. The plot displays clearly that even though there is no significant difference in the total ice abundances predicted by the two models (Fig.\,\ref{fig:L1544_density}), there are large variations in the distributions over the grain sizes depending on time and the medium density.

The distribution of ices over the grain populations tends to be a straightforward function of the grain size in model~M3 over the entire density range, that is, the ice abundances decrease with grain radius. We highlight two notable features: 1) For $\rm N_2$ at early times we recover the same trend evident in Fig.\,\ref{fig:M1vsM3_grains} where the ice is most abundant in bin~4 (note however that this trend does not appear at $n({\rm H_2}) \lesssim 10^4 \, \rm cm^{-3}$), and 2) The relative water ice abundance is heavily dependent on the medium density, in stark contrast to the results shown in Fig.\,\ref{fig:M1vsM3_grains}. This is caused by the grain temperature gradient across the core -- in the center where $T_{\rm dust}$ is low, CR desorption dominates over hydrogenation processes on large grains. Hence the water ice content is heavily concentrated on small grains, whereas the abundances of carbon-containing molecules tend to be relatively higher on large grains.

The abundance profiles are much more complicated at low medium densities in model~M4, where the differences in $T_{\rm dust}$ between the size bins strongly affects the efficiencies of chemical processes on the grain surfaces. Most of the $\rm N_2$ and CO ice tends to be incorporated in the ice in size bins~3~and~4, which is in stark contrast to the distribution in model~M3. Water ice, being the end product of hydrogenation processes and very strongly bound to the grain surface, does not exhibit similar complicated size-dependent abundance variations. The predictions of models~M3~and~M4 are in near perfect agreement at densities above $n({\rm H_2}) \sim 10^5 \, \rm cm^{-3}$, where the temperature difference between the smallest and the largest grains is small.

A notable result displayed in Fig.\,\ref{fig:L1544_density_ices} is that the relative ice abundances do not follow the relative abundances of the grains themselves. The general trend is that the smallest grains are underrepresented in this regard, that is, the relative ice abundance is lower than the corresponding grain abundance; in the other size bins, the trend is reversed (except for the largest grains at high density). The consequences of introducing a grain-size-variable equilibrium $T_{\rm dust}$ on the relative ice abundances are very interesting, because the effect is most pronounced in the density regime where ices start to form, in regions (partly) shielded from external UV radiation. The initial chemical abundances in our present model, in which the gas is initially (mostly) atomic, are not appropriate for the inner, high-density regions of the core model. However, our results do imply potential repercussions to the ice composition in the inner regions of pre-stellar cores as well, as the gas density increases through (quasi-static) contraction starting from lower-density gas. We test the effect of different initial conditions in Sect.\,\ref{ss:initialConditions}, where we also explore the role of the external radiation field on chemical evolution due to its effect on the equilibrium dust temperatures. Finally, we note that the thickness of the ice mantle that accumulates on the surface of a grain depends on the grain radius as well. We include for completeness some additional discussion on this topic in Appendix~\ref{appendixB}.

\section{Discussion}\label{s:discussion}

\begin{figure*}
	\includegraphics[width=2.0\columnwidth]{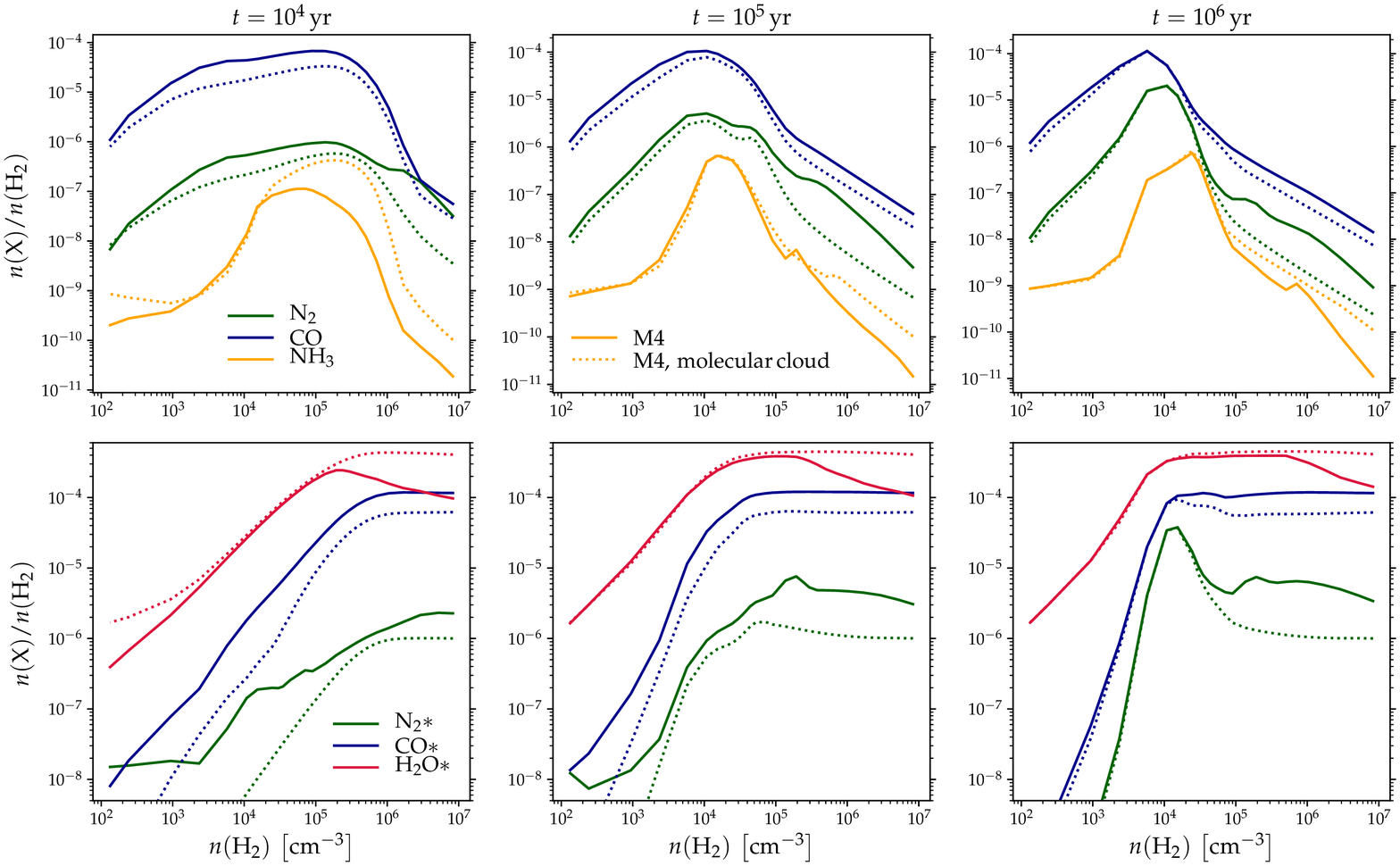}
    \caption{As Fig.\,\ref{fig:L1544_density}, but showing results from model~M4 (solid lines) and model~M4 with low-density cloud initial conditions (dotted lines). The time runs from the beginning of stage~2 (see main text).}
    \label{fig:L1544_density_molecularcloud}
\end{figure*}

\subsection{Effect of initial conditions}\label{ss:initialConditions}

We already pointed out above that the initial chemical composition adopted in this paper is not appropriate in the center of the core where the medium density is high ($n({\rm H_2}) > 10^6 \, \rm cm^{-3}$). Starting instead from an initially (mostly) molecular gas composition would influence the time-scales related to the development of ice chemistry, both for the total abundances and for the relative abundances on the different populations of grains. In a realistic scenario, the core structure is expected to develop gradually from a cloud with a lower average density, and chemical evolution will take place throughout the evolution of the physical structure. To emulate the chemical evolution during the core formation process, we constructed a simple two-stage model, where the chemistry is first let to evolve over some duration of time in a low-density environment roughly corresponding to molecular clouds, after which we extract the abundances of all species and use them as initial abundances for the core model. A similar process has been adopted in other works in the literature, for example to investigate the chemical properties of the starless core L1521E \citep{Nagy19}. For the molecular cloud properties, we took the values at the edge of the L1544 core model: $n({\rm H_2}) = 2.66\times10^2 \, \rm cm^{-3}$, visual extinction $A_{\rm V} = 1\,\rm mag$, $T_{\rm gas} = 18.7\,\rm K$, and $T_{\rm dust} = \left[13.3,13.5,13.7,13.9,13.9\right] \, \rm K$. We considered an array of grain temperatures, that is, we employed our model~M4 for this test. We ran the molecular cloud model for $t = 10^6 \, \rm yr$ (stage~1), and then ran the core model for an additional $t = 10^6 \, \rm yr$ (stage~2).

Figure~\ref{fig:L1544_density_molecularcloud} shows the results of these calculations for the standard model~M4 and its variant where chemical evolution starts from the abundances obtained at the end of stage~1. Given enough time for chemical evolution in the core phase (using the L1544 physical structure), the results in the two cases are virtually identical at low densities. Abundance differences of a factor of a few appear at densities above $n({\rm H_2}) \sim 10^5 \, \rm cm^{-3}$. Using molecular initial abundances does not have a clearly predictable effect on the results; switching the initial abundances causes some species to increase in abundance (e.g., water, ammonia), while others decrease in abundance (e.g., CO, $\rm N_2$). The results of the comparison are also affected by how long the gas is let to evolve in the molecular cloud. We have shown in \citet{Sipila19a} that running such a low-density model for $t = 10^5 \, \rm yr$ only does not have an appreciable effect on the chemistry in the core stage. The discussion is therefore affected by uncertainties involved in the time-scales of various processes, such as the duration of core formation and the lifetime in the pre-stellar phase. Abundance variations of a factor of a few across (a part of) the core, while seemingly small, may have a significant effect on observable line emission depending on critical densities of the lines \citep{Sipila18}.

\begin{figure}
	\includegraphics[width=1.0\columnwidth]{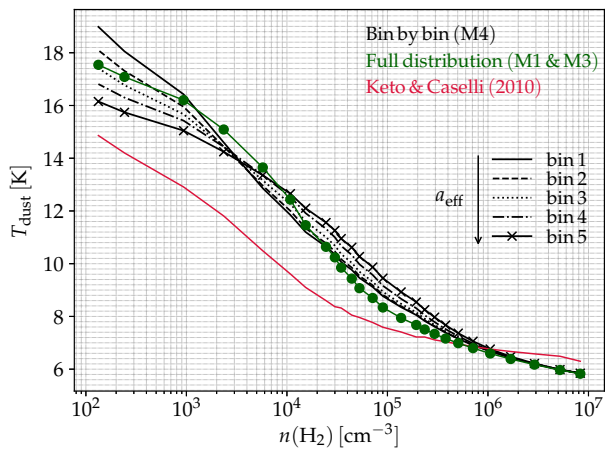}
    \caption{As Fig.\,\ref{fig:L1544_temperatures}, but the core is subjected to the unattenuated ISRF.}
    \label{fig:L1544_temperatures_AV0}
\end{figure}

Another property that has a crucial influence on our results in the outer core is the strength of the ISRF. If the field strength is increased, in particular in the UV, there are implications not only to the absolute dust temperatures but also to the relative dust temperatures in the size bins. We investigated this issue by running another variant of the M4~model where the core is subjected to the unattenuated ISRF (that is, $A_{\rm V}^{\rm ext} = 0 \, \rm mag$). Figure~\ref{fig:L1544_temperatures_AV0} shows the $T_{\rm dust}$ profiles obtained when omitting the visual extinction external to the core. In this case a turnover point appears at a density of a few times $10^3\,\rm cm^{-3}$; at densities lower than this, the grain equilibrium temperatures decrease with grain size. This is because the smallest grains absorb UV wavelengths efficiently (we return to this point in Appendix~\ref{appendixC}). For densities above a few times $n({\rm H_2}) \sim 10^4\,\rm cm^{-3}$ where UV photons do not penetrate, the smallest grains are the coolest, and the dust temperatures in the models with $A_{\rm V}^{\rm ext} = 0$ or 1\,mag are very close to each other.

\begin{figure*}
\centering
	\includegraphics[width=2.0\columnwidth]{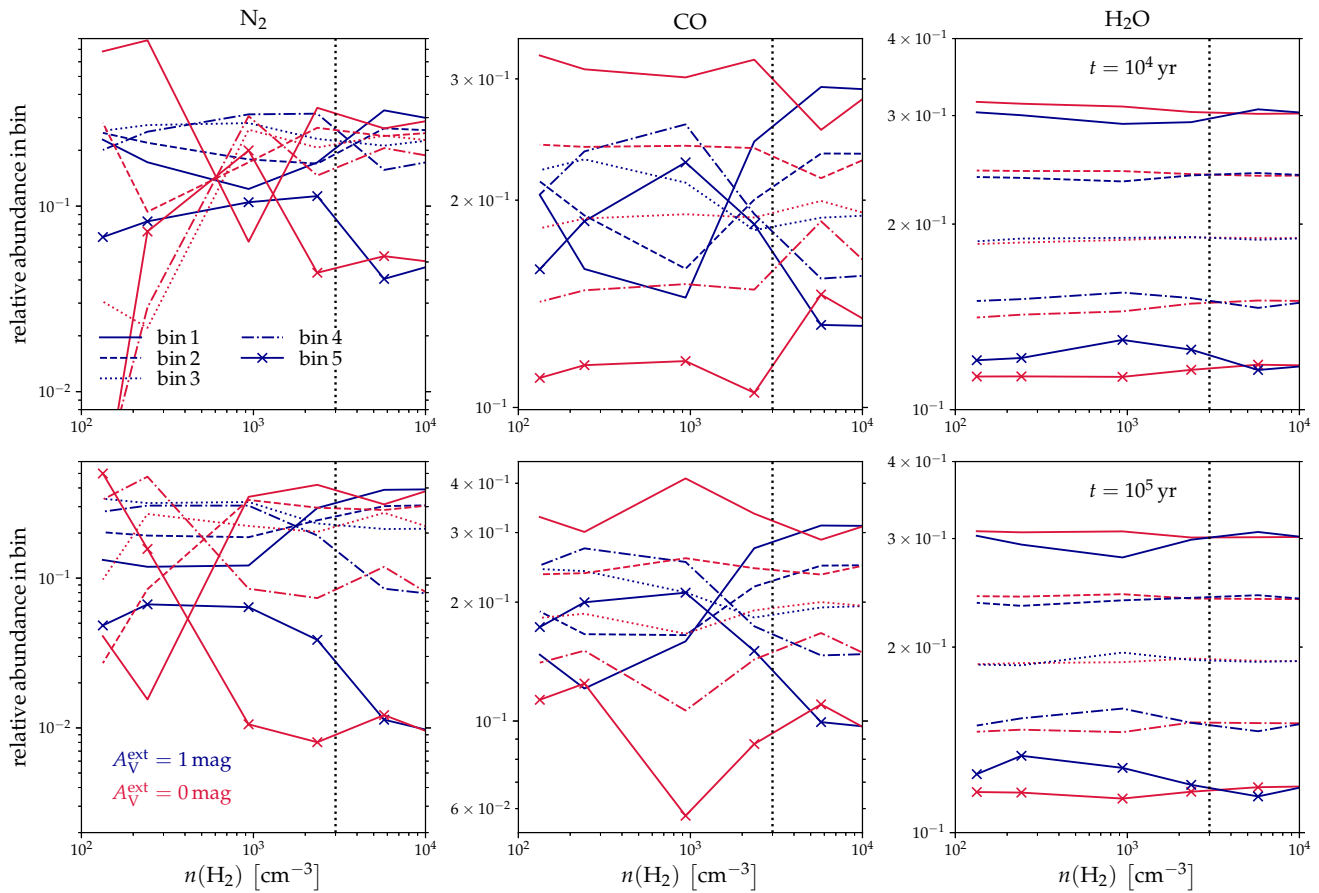}
    \caption{Relative abundances in each grain size bin as functions of density in the L1544 model at $t = 10^4 \, \rm yr$ ({\sl top}) and at $t = 10^5 \, \rm yr$ ({\sl bottom}) for model~M4. Blue lines correspond to $A_{\rm V}^{\rm ext} = 1\,\rm mag$, while red lines correspond to $A_{\rm V}^{\rm ext} = 0\,\rm mag$. The effective grain radius increases with bin label. Note the x-axis range which differs from that in Fig.\,\ref{fig:L1544_density_ices}. The dotted black line marks the approximate position of the turnover point of the dust temperatures (Fig.\,\ref{fig:L1544_temperatures_AV0}).}
    \label{fig:L1544_density_ices_Av0}
\end{figure*}

The modification in $T_{\rm dust}$ naturally affects molecular abundances as well. To quantify this issue, we ran model~M4 setting the dust temperature profiles to those shown in Fig.\,\ref{fig:L1544_temperatures_AV0}, {\sl but retaining the $A_{\rm V}^{\rm ext} = 1 \, \rm mag$ assumption in the chemical model}. This is an obvious inconsistency between the radiative transfer and chemical models, but it allows us to quantify more properly the effect of modifying $T_{\rm dust}$ alone, as moving from $A_{\rm V}^{\rm ext} = 1 \, \rm mag$ to 0\,mag would also translate to a significant increase in the efficiency of photoreactions, affecting the chemical evolution in the gas phase and in the ice. Figure~\ref{fig:L1544_density_ices_Av0} shows the relative ice abundances in the models with $A_{\rm V}^{\rm ext} = 1$ or $0 \, \rm mag$, concentrating on the low-density outer part of the core. Once again the response of the ice to changes in the grain properties depends on the molecule. For CO in the model with $A_{\rm V}^{\rm ext} = 0\,\rm mag$, the relative ice abundance distribution in each size bin is straightforward and, notably, CO ice is most abundant in bin~1 despite that grain population having the highest equilibrium temperature at low medium densities. $\rm N_2$ ice however shows again its great sensitivity to the physical conditions as far as its formation is concerned, while water ice does not respond strongly to the changes in $T_{\rm dust}$.

We reiterate that in these tests we neglected the effect of the one magnitude decrease in visual extinction on photochemistry, and that we would expect the (ice) chemistry to change drastically if $A_{\rm V}^{\rm ext}$ was decreased also in the chemical model. Furthermore, it must also be noted that: 1) Fig.\,\ref{fig:L1544_density_ices_Av0} shows the {\sl relative} ice abundances in each grain size bin, and that the total ice abundances are very low in these conditions, and 2) The high dust temperature in the outskirts of the core means that thermal desorption of relatively weakly bound species (such as atomic~H~and~N) is comparable to, or overpowers, CR desorption. Nevertheless our results clearly show that the ice composition can be strongly affected by the introduction of a grain size distribution, further underlining the demand for a detailed treatment of chemical evolution throughout the formation of a core, as opposed to assuming a static physical structure. Even such a detailed model is, unfortunately, affected by uncertainties from many sources, such as the initial elemental abundances.

\subsection{General trends in molecular abundances}\label{ss:otherAbundances}

\begin{figure*}
	\includegraphics[width=2.0\columnwidth]{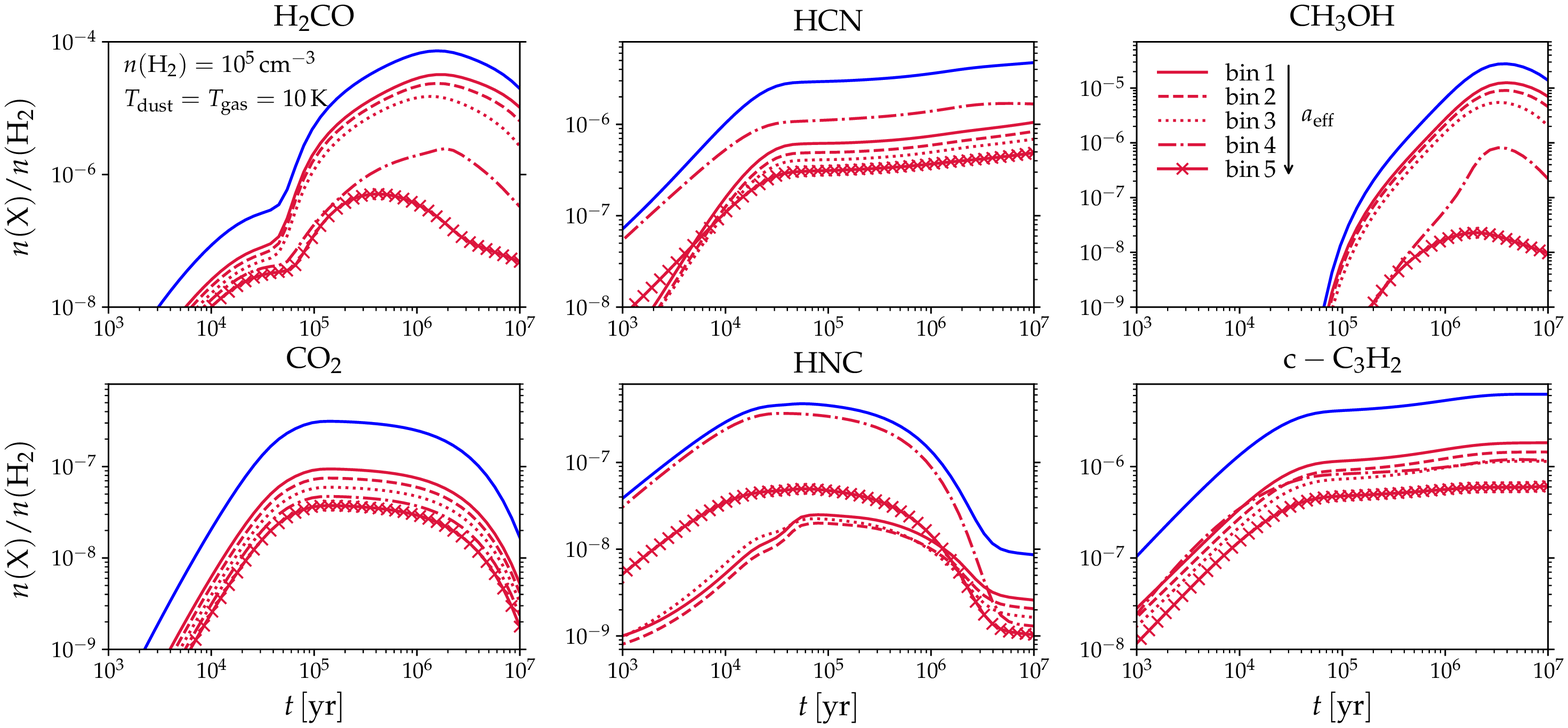}
    \caption{Distributions of selected molecules in the ice (indicated on top of each panel) as functions of time, in model~M3 at $n({\rm H_2}) = 10^5 \, \rm cm^{-3}$ and $T = 10\,\rm K$. The red lines represent the abundances on each grain population, while the blue line represents the sum over the individual grain populations. The effective grain radius increases with bin label as indicated in the plot.}
    \label{fig:M1vsM3_grains_additionalspecies}
\end{figure*}

In the above, we showed for $\rm N_2$ and CO ice that the abundance distributions on different grain populations in model~M3 are straightforward at late times: abundances decrease as the grain radius increases. This finding does not extend to all species. Fig.\,\ref{fig:M1vsM3_grains_additionalspecies} shows the distributions of selected additional molecules in model~M3. $\rm H_2CO$ and $\rm CH_3OH$ all present ice abundance profiles where the abundance decreases monotonously with increasing grain radius. However, HCN and HNC are both most abundant in bin~4 virtually throughout the time evolution, and in the case of HNC, the fifth size bin dominates over the smallest three over a long time period of time. $c-\rm C_3H_2$ is a mixed case as it presents an enhanced abundance in bin~4 over that in bin~3, even though the abundance in the other size bins decreases monotonously with increasing grain size. As explained above, these effects are tied to attributes such as binding energies and volatilities, and also the gas-phase/grain-surface origin of the species. No correlation to the chemical relationship between pairs of species exists. For example, $\rm H_2CO$ and $\rm CH_3OH$ (both produced mainly by the hydrogenation of CO) behave similarly to CO, but $\rm CO_2$ presents a distribution over the grain populations that is similar to that of $\rm H_2O$. This is in qualitative agreement with the connection between $\rm H_2O$ ice and $\rm CO_2$ ice that has been established observationally \citep[see for example][]{Boogert15}. However, we note that the fraction of $\rm CO_2$ ice that our model predicts (with respect to the most abundant ice constituent $\rm H_2O$) is very low, implying that a revision of $\rm CO_2$ formation in our model is needed in order to draw firm conclusions on this molecule.

Our results imply interesting consequences for the initial chemical composition of forming protostellar systems. This statement can be divided into two categories. First, introducing a distribution of grain sizes influences both gas-phase and ice abundances in high-density regions deep inside pre-stellar cores (also affecting the strength of the gas-dust collisional coupling; \citealt{Ivlev19}). This alone is a clear indication of the need to consider a size distribution as opposed to simply assuming that all grains have the same size. Second, the relative chemical composition of the ice on grains of different sizes is such that at high medium densities most of the molecular material is on the smallest grains. It is likely that the smallest grains are rapidly removed from the distribution owing to impacts with larger grains at high densities regulated by ambipolar diffusion (Silsbee et al., subm.) as the collapse progresses. The heat generated in such collision events may lead to the small grains shedding their mantles as they are heated to higher temperatures than the large grains. The efficiency of this process depends on the relative velocity of the colliding grains, and it may also be that other mechanisms in addition to ambipolar diffusion would be needed to enhance the desorption efficiency on small grains enough to impact gas-phase abundances. Regardless, the critical question is then related to the time-scales of molecular freeze-out and core collapse. If the collapse proceeds faster than the required time for the re-adsorption of previously evaporated molecules onto the (large) grains, the total amount of ice going into the prototellar phase would be smaller than in the case where no grain-grain collisions occur. Also, the ice content would be different than in the case of no collision (or otherwise)-induced desorption, owing to the fact that some species can be most abundant on large grains as opposed to small grains. This effect could conceivably account for the observed presence of gas-phase ammonia in the center of L1544 \citep[][]{Crapsi07} and, at least partly, for the observed abundance differentiation in a variety of molecules around its center \citep[][]{Spezzano17}. We will investigate this important issue in a future work.

The introduction of the grain size distribution affects the ionization fraction of the cloud, because the high amount of small grains leads to an overabundance of negatively ionized grains as compared to models considering monodisperse grains. We find that the size distribution models predict, for medium densities above a few $\times 10^4 \, \rm cm^{-3}$ where grain processes become important, a factor of a few less free electrons than model~M1 does, depending on the time. A more quantitative analysis is however challenging because of the way we have defined the size distribution -- we fixed the total grain area to correspond to that in the monodisperse grains model, which means that the overall number density of grains is higher in the size distribution model.

\subsection{Comparison to previous works \& restrictions of the present model}

There are, to our knowledge, two previous works in the literature that have discussed the effect of a grain size distribution on molecular abundances in physical conditions similar to those in the present paper, including grain-size-dependence for CR desorption in their model \citep{Zhao18, Iqbal18}. Here we briefly compare our present approach to these two papers. We note that other authors have studied the chemical effects arising from a grain size distribution without modifying the CR desorption efficiency as a function of grain radius \citep[e.g.,][]{Acharyya11,Pauly16}.

The CR desorption rate coefficients calculated in the present work are derived analogously to the approach of \citet{Zhao18}, and we have verified that for the same minimum and maximum grain size limits, and discretization of the distribution, we obtain the same duty cycles and transient maximum grain temperatures as \citet{Zhao18} did. However, there is a significant difference in the chemical models as they did not consider grain-surface chemical reactions at all (but allowed molecules to adsorp onto and desorb from grains). Therefore our present results are not comparable to theirs, and we do not for example recover the result from \citet{Zhao18} that the abundances of nitrogen carriers in the gas phase could be enhanced by the introduction of a grain size distribution. We did confirm through benchmarking calculations that if we disregard grain-surface reactions and use the same chemical network as \citet{Zhao18}, we obtain similar overall results (some differences related to input parameters remain).

\citet{Iqbal18} studied chemical abundances in physical conditions corresponding to cold cores by employing a gas-grain chemical model that includes a treatment of grain-size-dependence in CR desorption rate coefficients. They did not however account for the changes in the grain cooling timescale induced by the variation in the transient maximum grain temperature, and consequently the duty cycles and hence the CR desorption rate coefficients that they derived are very different from those presented here (and in \citealt{Zhao18}). Also, they did not fix the total grain surface area between the monodisperse and size distribution models as we do in this work, making our respective approaches incompatible. Nevertheless we obtain broadly similar results in some respects as \citet{Iqbal18} did, for example that gas-phase abundances in models with and without grain size distribution typically agree within an order of magnitude.

We adopted in this work the grain size distribution from \citet{Mathis77}, in which the grain abundance has a simple power-law dependence on the grain size. Switching to another size distribution model where the relative amounts of small and large grains are different from the present case would naturally affect the ice compositions on the various grain populations. However, as long as the total grain surface area was kept constant with respect to our fiducial monodisperse grains model, we would expect no significant effect on gas-phase abundances due to such a switch. We do not expect our conclusions to change in a fundamental way if the size-dependence of the distribution is varied, unless there is a strong overabundance of large grains due to grain growth.

Another property that affects our results, especially pertaining to the relative ice abundances, is the adopted optical properties of the grains. Here we constructed custom grain models mimicking the (unprocessed) thin ices case of \citet{Ossenkopf94}. The new models are assumed to apply at all densities in the model cloud, that is, we do not modify the grain optical properties as a function of chemical evolution. The differences in grain equilibrium temperature over the size distribution are the result of density-dependent attenuation of external radiation. Obviously, the assumption of thin ices on the grains, as far as optical properties are concerned, is inappropriate at low medium densities early into the cloud evolution where no significant ice formation has yet taken place. Conversely, at high densities the ices will quickly grow thicker than what is assumed in our opacity model. It is possible with our current model setup to modify the dust opacities time-dependently based on the ice thickness (separately in each grain size bin), simulating the real-time effect of evolving ice layers on the optical properties of the grains, as a function of position in the cloud. Such an undertaking is however beyond the scope of the present paper, and will be presented in a future work.

\section{Conclusions}\label{s:conclusions}

We investigated the effect of a grain size distribution on molecular abundances in physical conditions corresponding to starless and pre-stellar cores, using a gas-grain chemical model in which the cosmic ray (CR) desorption efficiency depends on the size of the grain. Small grains are heated to high temperatures by impinging CRs, but they also cool fast. Large grains cool slower and are heated by only a few tens of K by CRs. Consequently, the CR desorption rate coefficients are higher on large grains than on small grains. We divided the MRN size distribution into five distinct grain size bins and kept explicit track of the ice abundances in each bin as a function of time. The size distribution can be implemented in a chemical model in a straightforward manner, and we provided a table displaying the crucial parameters as a function of grain radius. We also considered a model where the grain equilibrium temperature is allowed to vary as a function of grain radius. For this, we determined the grain temperatures with a radiative transfer code using dust models custom-built for the present purposes.

We found that the grain-size-variable CR desorption efficiency has highly non-trivial consequences for the ice content. Some molecules are most abundant on the smallest grains (e.g., CO, water, $\rm CO_2$), while others can be most abundant on larger grains (e.g., HCN and HNC) -- although not necessarily on the largest grains in the distribution. This effect is tied to the main formation mechanism of a given molecule. CO for example is mainly produced in the gas phase, and its grain-surface abundance is a simple decreasing function of the grain size. $\rm N_2$ however is formed at early times almost exclusively on grain surfaces where the competition between desorption and hydrogenation influences the $\rm N_2$ formation efficiency as a function of grain size. As a consequence, the grain-surface $\rm N_2$ abundance is at early times highest on the larger grains in the distribution, but is later concentrated on small grains because of the high CR desorption efficiency on large grains. Some other molecules, such as HCN, can be most abundant on large grains over a very long period of time (even $> 10^6\,\rm yr$).

Allowing the equilibrium grain temperature to vary with grain radius introduces a temperature gradient of about one~K between the smallest and the largest grains, depending on the physical conditions. In regions where external UV radiation penetrates, the smallest grains are the warmest of the distribution, but the situation is reversed in the inner, well-shielded, regions inside starless and pre-stellar cores. The temperature gradient has a strong effect on the relative abundances of the ices on the different grain populations at low medium densities. However, allowing or disallowing variations in the equilibrium grain temperature does not have a large effect on the total abundances of grain-surface molecules or on gas-phase abundances; the practical difference is only in the relative abundances on the grain populations.

In the present paper we only considered cases where there is no evolution of the physical conditions from low density to high density through gravitational contraction. The strong influence of the variations in equilibrium grain temperature on the ice composition at low medium density, where ices begin to form, implies that the ice content at high density could be influenced as well as the parental cloud contracts and a core forms. Another fundamental finding in the present work is that the distribution of molecules on grains of different size is not uniform. We propose that grain-grain collisions in infalling gas (possibly boosted by ambipolar diffusion) that result in grain coagulation and the (partial) evaporation of the ice mantles on small grains, could contribute to the observed chemical differentiation toward the centers of starless and pre-stellar cores, for example in L1544 \citep[][]{Spezzano17}, L183 \citep{Swade89,Lattanzi20}, and TMC-1 \citep{Pratap97}. These findings imply great consequences for the chemical composition in star-forming clouds, and so in a future work we will investigate the time-evolution of ices in a collapsing core, and will also explore the evaporation of ice mantles on small grains due to grain-grain collisions.

\begin{acknowledgements}
The authors thank the referee, Dr. Tyler Pauly, for constructive comments on the manuscript that helped to improve the presentation. The support by the Max Planck Society is gratefully acknowledged. O.S. thanks Mika Juvela and Mika Saajasto for helpful discussions related to the grain models.
\end{acknowledgements}

\bibliographystyle{aa}
\bibliography{sizeDistribution.bib}

\appendix

\onecolumn

\section{Variations in the number of grain size bins}\label{appendixA}

\begin{figure*}
	\includegraphics[width=1.0\columnwidth]{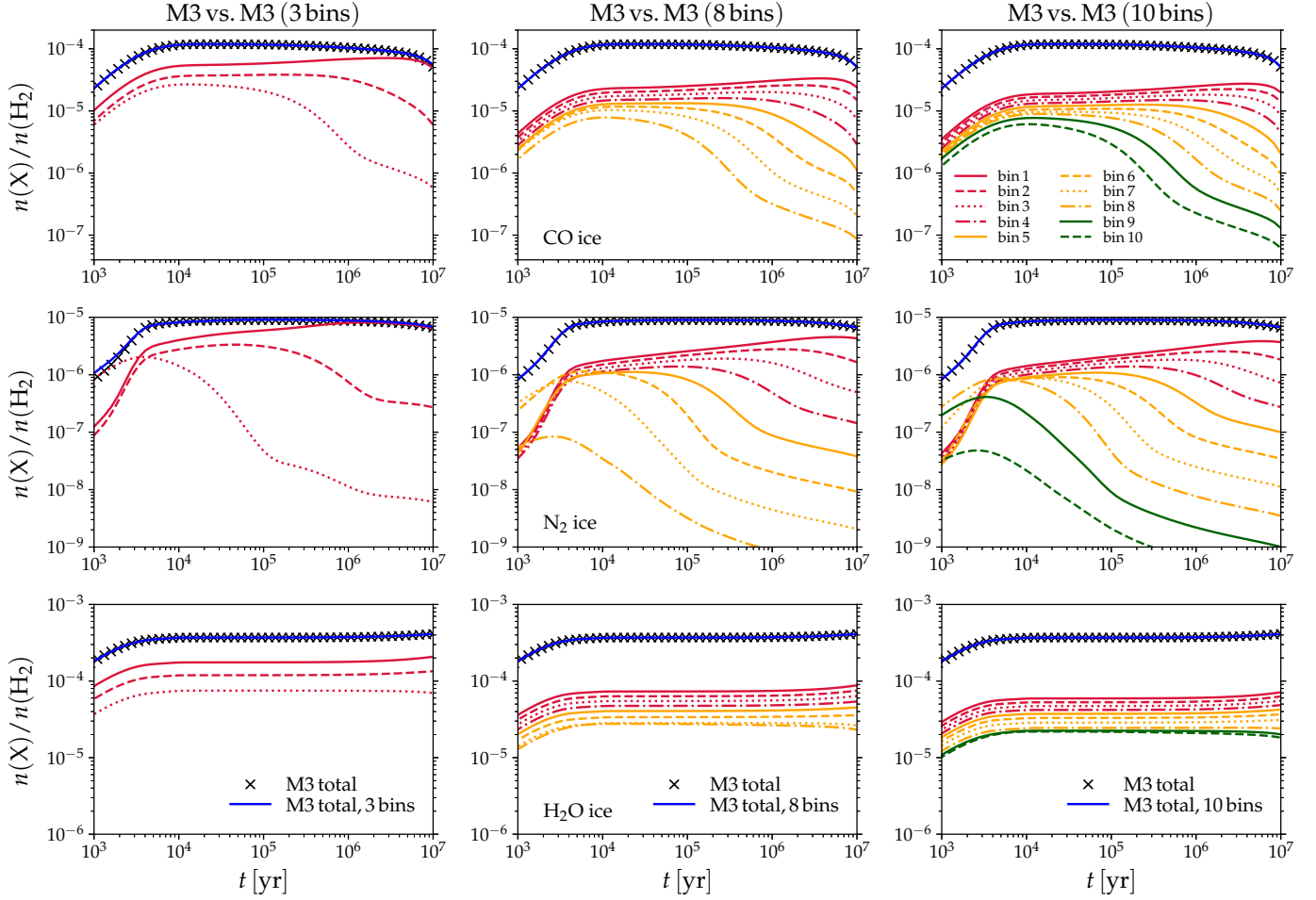}
    \caption{Distributions of CO ({\sl top row}), $\rm N_2$ ({\sl middle row}), and $\rm H_2O$ ({\sl bottom row}) ice as functions of time at a fixed medium density of $n({\rm H_2}) = 10^6 \, \rm cm^{-3}$. The left-hand, middle and right-hand panels show the results for model~M3 using 3, 8, or 10 grain size bins, respectively. Blue solid lines indicate the total ice abundance, while black crosses indicate the corresponding total abundance in our fiducial M3 model. Red, orange, and green lines represent the breakdown of the total ice abundance in the grain size bins as displayed in the top right panel. This breakdown is omitted for the fiducial M3 model for clarity.}
    \label{fig:M1vsM3_grains_binNumberTest}
\end{figure*}

The choice of the number of discrete grain sizes naturally affects ice abundances. When the number of grain size bins is increased, the material adsorbed from the gas phase is spread out over an increasing number of grains. However, as long as the limits of the size distribution (minimum and maximum grain radius) are kept fixed along with the other critical parameters such as the dust-to-gas mass ratio, the chosen number of size bins does not affect the total grain surface area, and hence should not have an effect on our results at least when the bin number is greater than a few.

We compare in Fig.\,\ref{fig:M1vsM3_grains_binNumberTest} the ice abundances in our fiducial model~M3 (five grain size bins) against variants of model~M3 where the grain size bin number is either three, eight, or ten. In the model with the fewest size bins, some very small deviation from the fiducial model is found at early times, but none is apparent after $t \gtrsim 10^4 \, \rm yr$. Increasing the number of size bins decreases the abundances on the individual grain populations, but the total abundance (sum over all grain populations) is unaffected by the choice of the number of grain sizes. We also recover similar chemical behavior for the ices as in the fiducial case, for example, that the $\rm N_2$ ice abundance is at early times dominated by the larger grains in the distribution. The abundances of gas-phase molecules are similarly unaffected by the number of grain size bins. These results justify our choice of five size bins as being enough to obtain accurate results while simultaneously limiting the required computational time, which increases non-linearly with each added grain size bin.

\section{Ice thickness as a function of grain size}\label{appendixB}

\begin{figure}
\centering
	\includegraphics[width=1.0\columnwidth]{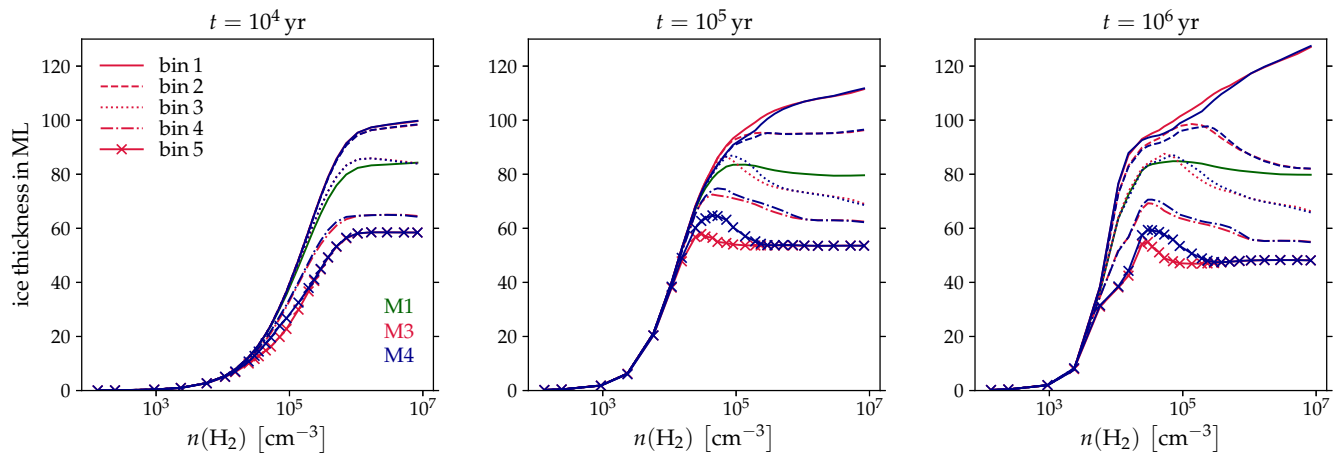}
    \caption{Ice thickness in monolayers (MLs) as a function of density in L1544 at three different times. The green, red, and blue lines correspond to models~M1, M3, and M4, respectively. The grain size increases with bin label.}
    \label{fig:L1544_icethickness}
\end{figure}

In the main text, we have in several instances emphasized that the ice abundances are the largest on the smallest grains, which is sensible given that they represent the largest portion of the total grain surface. Because the number of binding sites on the grain surface decreases with grain radius, one expects an accompanying decrease in the ice thickness. Figure~\ref{fig:L1544_icethickness} shows the ice thickness in monolayers (ML) for models~M1, M3, and M4, as a function of density in the L1544 core model. We highlight three features of the plot. First, the ice accumulates very slowly at low density; for example it takes $\sim 10^6$\,yr to grow just two MLs of ice at $n({\rm H_2}) = 10^3 \, \rm cm^{-3}$. Second, in the monodisperse grains model (M1), the ice thickness does not evolve much after $t = 10^5$\,yr, and peaks around $n({\rm H_2}) = 10^5 \, \rm cm^{-3}$ because the CR desorption efficiency increases with density, being especially important at the highest densities. In models M3~and~M4, a similar peak is evident for all but the smallest grains, for which the ice thickness keeps increasing with density owing to the inefficiency of CR desorption on small grains. For the other grain sizes, the peak shifts to lower densities as the grain size increases, in accordance with the increase in CR desorption efficiency. There is no significant difference between the ice thickness in models~M3~and~M4. Third, the ice thickness is a simple decreasing function of grain size at all densities. This means that the high variability of the relative ice content seen for example in Fig.\,\ref{fig:L1544_density_ices}, where for example $\rm N_2$ ice is at times more abundant on large grains than on small grains, is not due to the ices accumulating at varying rates.

\section{Grain-size-dependent optical depth}\label{appendixC}

\begin{figure}
\centering
	\includegraphics[width=0.8\columnwidth]{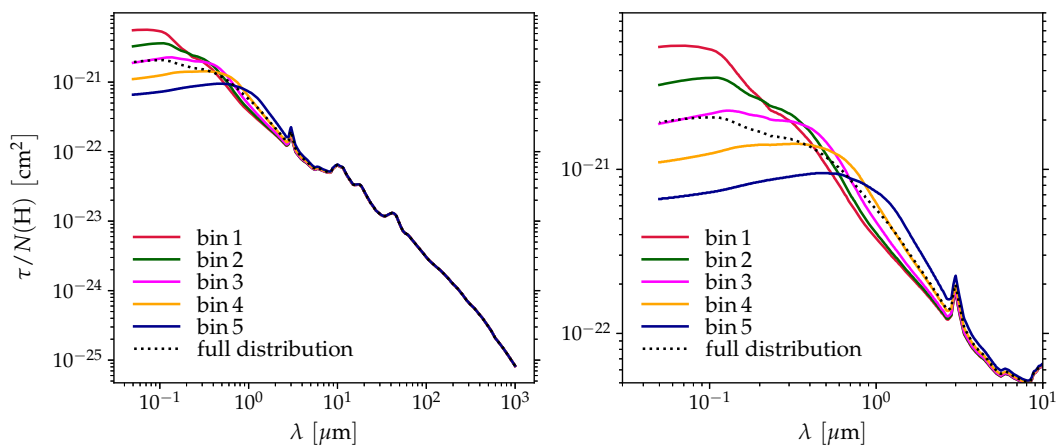}
    \caption{Optical depth per hydrogen column density for the six grain models constructed for this paper, that is, the five models corresponding to the size bins and the one calculated over the entire size distribution. The grain size increases with bin label. The right-hand panel shows a zoom-in of the short-wavelength part of the spectrum.}
    \label{fig:extinctionCurves}
\end{figure}

We tested in Sect.\,\ref{ss:initialConditions} the effect of extinction on the size-dependent equilibrium dust temperatures. Here, we complement that discussion by showing in Fig.\,\ref{fig:extinctionCurves} the optical depth per hydrogen column density for the six grain models constructed for this paper using SIGMA. The plot serves to explain the density-dependence in the equilibrium grain temperatures shown in Figs.\,\ref{fig:L1544_temperatures}~and~\ref{fig:L1544_temperatures_AV0} in the main text. In our fiducial model where $A_{\rm V}^{\rm ext} = 1\,\rm mag$, the UV part of the ISRF spectrum is strongly attenuated already at the core edge, and the grain temperatures are dictated by the optical properties of the grains at longer wavelengths. The optical depth in the near-to-mid infrared wavelength range is a simple decreasing function of grain size. The equilibrium grain temperatures are thus highest for the largest grains when there is appreciable extinction external to the core. However, when this extinction is removed from the model, the grain temperatures in the outer core are dominated by the optical properties in the UV part of the spectrum, and consequently the smallest grains are then the warmest.

\end{document}